\documentclass[3p,sort&compress,times,letterpaper]{elsarticle}
\usepackage[T1]{fontenc}
\usepackage[latin9]{inputenc}
\usepackage{float}
\usepackage{amsbsy}
\usepackage{amsmath}
\usepackage{amstext}
\usepackage{hyperref}
\usepackage{graphicx}
\usepackage{esint}
\usepackage{subfig}

\begin{document}

\begin{frontmatter}

\title{Topological susceptibility near $T_{c}$ in SU(3) gauge theory}


\author{Guang-Yi Xiong$^{a}$, Jian-Bo Zhang$^{a}$}
\address{$^{a}$Department of Physics, Zhejiang University, Zhejiang 310027,
P.R. China}

\author{Ying Chen$^{b,c}$ }
\address{$^{b}$Institute of High Energy Physics, Chinese Academy of Sciences,
Beijing 100049, P.R. China\\$^{c}$Theoretical Center for Science Facilities, Chinese Academy
of Sciences, Beijing 100049,P.R. China}

\author{Chuan Liu$^{d,e}$}

\address{$^{d}$School of Physics, Peking University, Beijing 100871, P.R. China}
\address{$^{e}$Collaborative Innovation Center of Quantum Matter, Beijing 100871, P.R. China}

\author{Yu-Bin Liu$^{f}$}
\address{$^{f}$School of Physics, Nankai University, Tianjin 300071, P.R.
China}

\author{Jian-Ping Ma$^{g}$}
\address{$^{g}$Institute of Theoretical Physics, Chinese Academy of Sciences,
Beijing 100080, P.R. China}

\begin{abstract}
Topological charge susceptibility $\chi_{t}$ for pure gauge SU(3) theory at finite temperature is studied using anisotropic lattices. The over-improved stout-link smoothing method is utilized to calculate the topological charge. Near the phase transition point we find a rapid declining behavior for $\chi_{t}$ with values decreasing from $(188(1)\mathrm{MeV})^{4}$ to $(67(3)\mathrm{MeV})^{4}$ as the temperature increased from zero temperature to $1.9T_{c}$ which demonstrates the existence of topological excitations far above $T_{c}$. The 4th order cumulant $c_4$ of topological charge, as well as the ratio $c_4/\chi_t$ are also investigated. Results of $c_4$ show step-like behavior near $T_c$ while the ratio at high temperature agrees with the value as predicted by the diluted instanton gas model.
\end{abstract}


\end{frontmatter}


\section{Introduction}

Quantum Chromodynamics (QCD) is the gauge field theory that describes the rules of strong interaction among quarks and gluons, which has achieved great success in describing modern particle experiments. Higgs boson--used to be the last missing particle in this framework -- was found in 2012 on LHC, claiming the landmark completeness of the Standard Model. From 2000s, lattice QCD which simulates QCD on finite lattices by powerful computers, has gained enormous amount of results in accordance with experiments on accelerators, thanking for the incredible development of computers as well as the continuous progress in computational methods. Despite of the success in the spectroscopy of hadrons, lattice QCD researchers have made great efforts on QCD thermodynamics with nonzero temperature and chemical potential, from which further understanding about hadronic matters and quark-gluon plasma can be obtained and problems such as the physics in early universe with extremely high temperature and density can be addressed.

Topology in lattice QCD has been widely studied for various purposes, while one of the most exciting applications is to explore the structure of QCD vacuum, which has not been well understood. The topological susceptibility $\chi_t$ has attracted special interest since 1979, when the quenched $\chi_{t}$ was related to the $\mathrm{U}\left(1\right)$ axial anomaly and mass of $\eta'$ meson through the well-known Witten-Vaneziano relation~\cite{witten1979current,veneziano1979u}. Furthermore, the distribution of the topological charge $Q$ can be described in terms of its cumulants~\cite{reichl1980modern}:
\begin{eqnarray}
c_{2n}&=&\left.\frac{d^{2n}e_{vac}\left(\theta\right)}{d\theta^{2n}}\right|_{\theta=0},\\
c_{2}&=&\chi_{t}=\left.\frac{\left\langle Q^{2}\right\rangle }{V}\right|_{\theta=0},\label{eq:chi}\\
c_{2n}&=&\left(-1\right)^{n+1}\left[\frac{\left\langle Q^{2n}\right\rangle }{V}+\sum_{m=1}^{n-1}\left(-1\right)^{m}\left(\begin{array}{c}
2n-1\\
2m-1
\end{array}\right)\left\langle Q^{2\left(n-m\right)}\right\rangle c_{2m}\right],~~n\ge2\label{eq:c2n}
\end{eqnarray}
where $e_{vac}\left(\theta\right)$ is vacuum energy density in the $\theta$ vacuum. The 2nd order cumulant $c_2$ is known as the topological susceptibility $\chi_{t}$, and $c_{4}$ -- the 4th order cumulant of $Q$ is important to lattice calculations of observables with fixed topology sector~\cite{Brower:2003yx,Aoki:2007ka}. Nonzero values for $c_{2n}$ with $n>1$ indicates non-Gaussian distribution of $Q$~\cite{Giusti:2007tu,Ce:2015qha}. From (\ref{eq:chi}) and (\ref{eq:c2n}), it is easy to see that $c_{4}=-\left(\left\langle Q^{4}\right\rangle -3\left\langle Q^{2}\right\rangle ^{2}\right)/V$.

On the other hand, lattice studies of QCD at finite temperature and the phase structure of QCD have made great progress recently~\cite{Petreczky2012}. In particular, the relation between chiral symmetry breaking and deconfinement transition, as both of them happen to be near $T_{c}$, is of great importance and has attracted many researchers. Since $\chi_{t}$ describes the topological fluctuations of the vacuum in quenched situation, the behavior of $\chi_t$ near $T_{c}$ is expected to provide a further understanding on the relation between chiral symmetry breaking and confinement.

Topological susceptibility at zero temperature for SU(3) gauge theory has been determined as $(191(5)\mathrm{MeV})^{4}$~\cite{DelDebbio2005}. $\chi_{t}$ at finite temperature also attracts great interests, especially around the transition temperature $T_{c}$~\cite{Alles:1996nm,deForcrand:1997ut,Gattringer2002,DelDebbio:2004rw}. In 2002, Gattringer {\em et al}~\cite{Gattringer2002} found the cross-over behavior of $\chi_{t}$ in the temperature interval of $0.8T_{c}\sim1.3T_{c}$, which is in accordance with results of other researchers. In their work chirally improved fermion action and fermionic method based on Atiyah-Singer index theorem were employed to calculate the topological charge. There are basically two kinds of methods to extract the topological charge: fermionic and bosonic method. For bosonic method, some kind of smoothing procedure always be applied first to dampen the fluctuations at the UV scale, while hopefully leaving the long range physics unchanged. In 2010 L{\"u}scher proposed the Wilson flow method~\cite{Luscher:2010iy} that could be employed to calculate topological charge, which is renormalizable and converges to the continuum definition. More recently, Glaudio Bonati et al studied the dependence of 4D SU(N) gauge theories on the topological term at finite temperature by using cooling method~\cite{Bonati:2013tt}, found that the $\theta$ dependence are drastically changed across the deconfinement transition. Moreover, the comparison between standard cooling method and Wilson flow method were also studied in Ref.~\cite{Bonati:2014tqa}, which led to equivalent results, both for average quantities and configuration by configuration. In this paper we present our work using over-improved stout-link smearing method~\cite{Moran2008} on the anisotropic lattices with a wider range of temperature to provide an alternative study on this interesting subject. This smearing method is based on the classical instanton solution in continuum limit, which is model dependent. However, it is observed that the method keeps the topological charge stable and always values around the same integer obtained by the fermionic method. This method is also relatively cheaper than the other smoothing method. More recent studies were also carried out in Ref.~\cite{Cossu:2012gm,Bornyakov:2013iva,Ilgenfritz:2013oda} and part of their results are compatible with ours.

The diluted instanton gas model gives the ratio of $c_{4}$ and $c_{2}$ as the value $-1$~\cite{Gross:1980br}. Recent calculation at zero temperature gives the ratio $R=-0.233(45)$~ \cite{Ce:2015qha} (different definition of $R$ leads to the opposite sign with ours in that paper), which is in contradiction with the value predicted by the diluted instanton gas model. At high temperature phase($>1.15T_{c}$) the domination of the diluted instanton gas model was confirmed through the calculation of $b_2$ and free energy~\cite{Bonati:2013tt}, where $b_2=c_4/12\chi_t$. In this paper, we calculate the ratio $R=c_4/\chi_t$ below and above $T_{c}$ with pure gauge configurations, trying to investigate the behavior of $R$ around $T_{c}$ using gluonic method.

This paper is organized as follows: we first present our lattice setup and details about extracting the topological charge $Q$ at finite temperature. We also discuss the over-improved stout-link method to smooth the gauge field and the calculation of $Q$ by field theoretical definition employing highly improved field strength tensor~\cite{BilsonThompson:2002jk}. After that, results of $Q$ and the susceptibility $\chi_t$, as well as the 4th order cumulant $c_4$ of topological charge and the ratio $R$ are presented and discussed. Finally main results of this paper are summarized.

\section{Lattice setting and method}

The calculation is carried out using anisotropic lattices, which have advantages of improving accuracy in lattice QCD for both zero and finite temperatures~\cite{Meng2009}. The Symanzik and tadpole improvement schemes of the gauge action are found to have better continuum extrapolation behaviors for many physical quantities, that is, the finite lattice spacing effect is well suppressed by these improvements. Considering these facts, we use the following improved gauge action, \cite{Morningstar:1997ff,Morningstar:1999rf,PhysRevD.73.014516}
\begin{equation}
S_{IA}=\beta\left\{ \frac{5}{3}\frac{\Omega_{sp}}{\xi u_{s}^{4}}+\frac{4}{3}\frac{\xi\Omega_{tp}}{u_{t}^{2}u_{s}^{2}}-\frac{1}{12}\frac{\Omega_{sr}}{\xi u_{s}^{6}}-\frac{1}{12}\frac{\xi\Omega_{str}}{u_{s}^{4}u_{t}^{2}}\right\}
\end{equation}
where $\beta$ is related to the bare QCD coupling constant $g_0$, $\xi=a_{s}/a_{t}$ is the aspect ratio for anisotropy (we take $\xi=5$ in this work), $u_{s}$ and $u_{t}$ are the tadpole improvement parameters of spatial and temporal gauge links respectively. $\Omega_{C}=\sum_{C}\frac{1}{3}\mathrm{ReTr}\left(1-W_{C}\right)$, with $W_{C}$ referring to the path ordered product of link variables along a closed contour $C$ on the lattice. $\Omega_{\mathrm{sp}}$ includes the sum over all spatial plaquettes on the lattice, $\Omega_{\mathrm{tp}}$ includes the temporal plaquettes, $\Omega_{\mathrm{sr}}$ includes the product of link variables about planar $2\times1$ spatial rectangular  loops, and $\Omega_{\mathrm{str }}$ refers to the short temporal rectangles (one temporal link, two spatial). Practically, $u_{t}$ is set to 1, and $u_{s}$ is defined by the expectation value of the spatial plaquette $P_{ss^{\prime}}$: $u_{s}=\left\langle \frac{1}{3}\mathrm{Tr}P_{ss^{\prime}}\right\rangle ^{1/4}$. Besides, it has been shown in Ref. \cite{Liu:2006dp} that the renormalization effects of the anisotropy $\xi$ is ignorable as $\beta$ varies. We also have used Wilson flow method to calculate the renormalized
anisotropy according to Ref. \cite{Borsanyi:2012zr} at different temperature with $\beta=3.2$, and the results show that the differences between the renormalized $\xi$ and bare ones are always less than $2\%$. We adopt anisotropic lattices with smaller $a_t$ compared to $a_s$, so that we can investigate higher temperature with small spatial lattice size.

\subsection{The definition of temperature }

The temperature on lattice is defined as follow:
\begin{equation}
T=\frac{1}{N_{t}a_{t}},
\end{equation}
where $N_{t}$ is the temporal lattice size. $T$ can be changed by varying either $N_{t}$ or the coupling constant $\beta$, which is related directly to the lattice spacing $a_t$. The critical temperature is determined for a given lattice $N_{t}=24$ after the critical coupling $\beta_{c}$ has been determined. The order parameter for determining $\beta_{c}$ is chosen to be the susceptibility $\chi_{P}$ of the Polyakov line, which is defined as
\begin{equation}
\chi_{_{P}}=\langle\Theta^{2}\rangle-\langle\Theta\rangle^{2},
\end{equation}
where $\Theta$ is the $Z_3$ rotated Polyakov line defined via:
\begin{equation}
\Theta=\begin{cases}
\text{Re}P\exp\left[-2\pi i/3\right] & \arg P\in[\pi/3,\pi)\\
\text{Re}P & \arg P\in[-\pi/3,\pi/3)\\
\text{Re}P\exp\left[2\pi i/3\right] & \arg P\in[-\pi,-\pi/3)
\end{cases}
\end{equation}
and $P$ represents the trace of the spatially averaged Polyakov line for each gauge configuration. After a rough scan, a more refined study for the peak position of the susceptibility $\chi_{_P}$ gives the critical coupling constant $\beta_{c}=2.808$, which corresponds to the critical temperature $T_{c}\approx0.724r_{0}^{-1}=296\mathrm{MeV}$~\cite{Liu:2006dp}. Here $r_{0}$ is the hadronic scale parameter and we take $r_{0}^{-1}=410\left(20\right)\mathrm{MeV}$.

Considering both finite volume effects and good resolution of temporal lattice at $T\sim2T_{c}$, we set $\beta=3.2$ in the study of topological susceptibility at finite temperature. The corresponding lattice spacing $a_s$ is obtained by calculating the static quark potential $V\left(r\right)$ on an anisotropic lattice $24^{3}\times128$. The fitting result of string tension $\sigma$ reads:
\begin{equation}
\frac{a_{s}}{r_{0}}=\sqrt{\frac{\sigma a_{s}^{2}}{1.6+e_{c}}}=0.1825\left(7\right),
\end{equation}
taking $r_{0}^{-1}=410\left(20\right)\mathrm{MeV}$, we have $a_{s}=0.0878\left(4\right)\mathrm{fm}$. Comparing to $a_{s}$ at $\beta_{c}=2.808$ and $N_{t}=24$, the approximate values for $N_{t}$ that corresponding to $T_{c}$ and
$2T_{c}$ at fixed $\beta=3.2$ can be calculated and the results are $N_{t}\sim38$ and $N_{t}\sim19$ respectively. More parameter details can be found in Ref.~\cite{Meng2009}. It should be pointed out that the difference between $T_{c}\left(\beta=2.808\right)$ and $T_{c}\left(\beta=3.2\right)$ is negligible due to the application of the improved gauge action.

In this study we change the temperature by varying temporal lattice size $N_t$. Working on the same $\beta$ which corresponding to the same $a_s$ makes all our lattices share the same spatial volume while the temperature $T$ varies in a wide range. This avoids the influence from possible finite volume effects arising from different spatial volumes. Setting $\beta=3.2$, we generated a series of lattice $24^{3}\times N_{t}$ with $N_{t}=$20, 24, 28, 32, 36, 40, 44, 48, 60, 80 and 128, which cover the range of $T\in\left[0.3T_{c},~1.9T_{c}\right]$, and the spatial size $L_s\approx2.1\mathrm{fm}$. For each anisotropic lattice with a fixed value of $N_{t}$, we sampled roughly 1000 configurations(1900 for $N_t=60$), after 10000 sweeps from cold start and with 500 sweeps between each sampling. Here one sweep consists of a composite update of 1 pseudo heat-bath and 5 over-relaxation procedures over all link variables. 5000 samples of $24^4$ isotropic lattice with the same $a_s$ were also generated and measured as the results at zero temperature. We roughly estimated the integrated autocorrelation time of topological charge $\tau_{Q}$, which is about 200 sweeps at zero temperature and about 440 sweeps at $1.9T_c$.

\subsection{Over-improved stout-link method}
It is well-known that the topological charge $Q$ calculated directly from a typical lattice configuration by the field theoretic definition~\cite{belavin1975pseudoparticle,luscher1982topology}:
\begin{equation}
Q=\frac{1}{32\pi^2}\int d^{4}x~\mathrm{tr}\left(F_{\mu\nu}\tilde{F}_{\mu\nu}\right),
\end{equation}
\begin{equation}
F_{\mu\nu}=\partial_{\mu}A_{\nu}-\partial_{\nu}A_{\mu}+ig\left[A_{\mu},A_{\nu}\right],~~~
\tilde{F}_{\mu\nu}=\frac{1}{2}\epsilon_{\mu\nu\rho\sigma}F_{\rho\sigma},
\end{equation}
will not be an integer in general, which is obviously conflicting with the continuum situation. Thus, some cooling or smearing procedure is essentially to be taken on original configurations before calculating $Q$ by bosonic methods, aiming to suppress the ultraviolet fluctuations. Highly improved lattice field-strength tensor\cite{BilsonThompson:2002jk} is also helpful to obtain an integer $Q$ on the lattice, as we used $1\times1$, $2\times2$ and $3\times3$ 3-loop improved $F_{\mu\nu}$ which is practically good enough.

All cooling or smearing methods are based on an approximation to the continuum gauge field action:
\begin{equation}
S_{g}=\frac{1}{2}\int d^{4}x~\mathrm{tr}\left[F_{\mu\nu}F_{\mu\nu}\right],
\end{equation}
and the lattice version of $S_g$ contains combination of plaquette and larger Wilson loops. However, the discretization error always exists, which is harmful to the topological objects and smoothing procedure. So in these methods the destruction of topological structure is unavoidable. For the purpose of the study on topology susceptibility, we adopt the over-improved stout-link smearing method developed by Moran and Leinweber~\cite{Moran2008} to filter the lattice, which is proved to be efficient in preserving instanton-like objects.

Over-improved stout-link smearing method introduces the over-improved parameter $\epsilon$~\cite{GarciaPerez:1993ki} into stout-link smearing algorithm~\cite{Morningstar:1997ff}. The improved action with instanton solution substituted reads:
\begin{equation}
S^{inst}\left(\epsilon\right)=\frac{8\pi^{2}}{g^{2}}\left[1-\frac{\epsilon}{5}\left(\frac{a}{\rho_{inst}}\right)^{2}+\frac{14\epsilon-17}{210}\left(\frac{a}{\rho_{inst}}\right)^{4}\right],
\end{equation}
where $\rho_{inst}$ is the instanton size. Negative $\epsilon$ leads to a positive leading order error term of $\rho_{inst}^{-2}$, which means instanton-like objects will be preserved when the action is reduced in smoothing procedure. The modified link combination used by over-improved stout-link method is as follow:
\begin{eqnarray}
\mathcal{C}_{\mu}\left(x\right) & = & \rho_{\mathrm{sm}} \sum\bigg\{\frac{5-2\epsilon}{3}\left(1\times1~ \mathrm{paths~touching~ }U_{\mu}\left(x\right)\right)\nonumber \\
 &  & -\frac{1-\epsilon}{12}\big(1\times2 + 2\times1\mathrm{~paths~touching~}U_{\mu}\left(x\right)\big)\bigg\}.
\end{eqnarray}

\begin{figure}[h t b]
\begin{centering}
\includegraphics[scale=0.5]{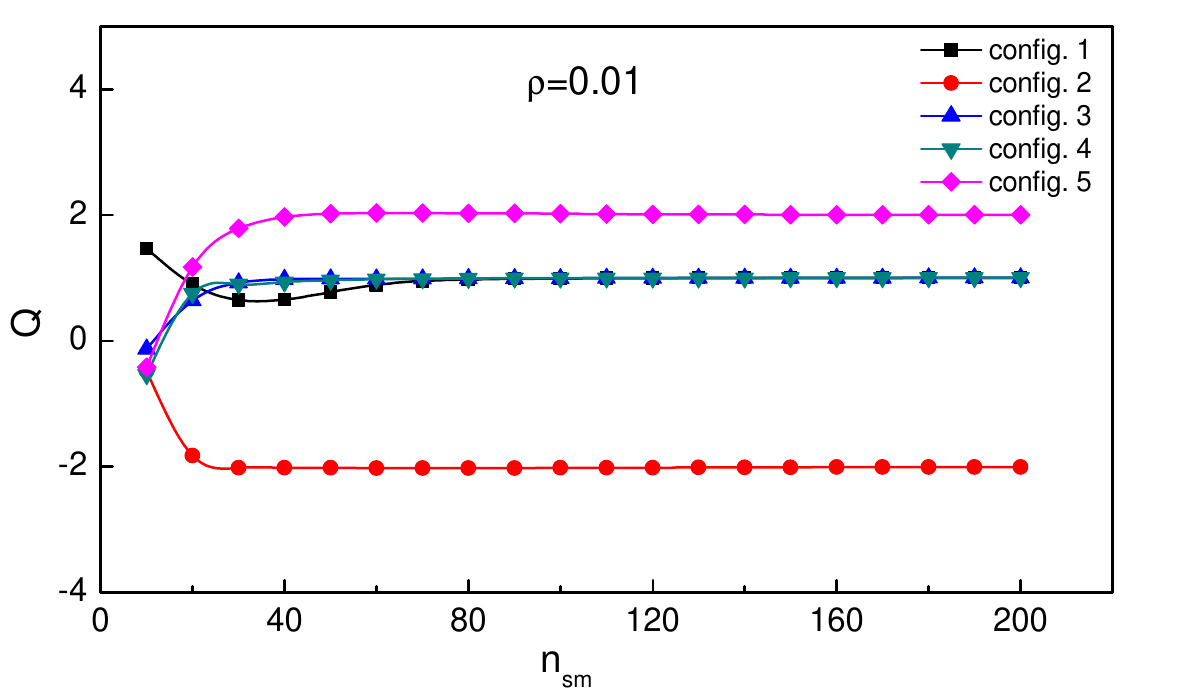}$~~~~$ \includegraphics[scale=0.5]{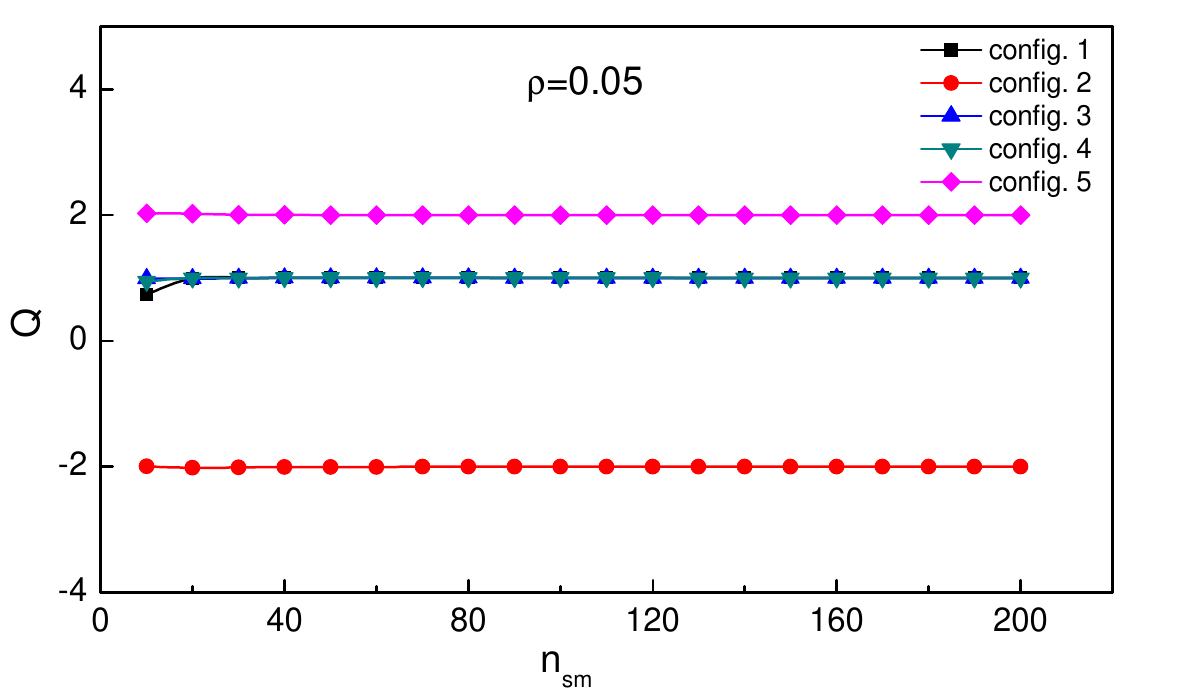}
\par\end{centering}
\begin{centering}
\includegraphics[scale=0.5]{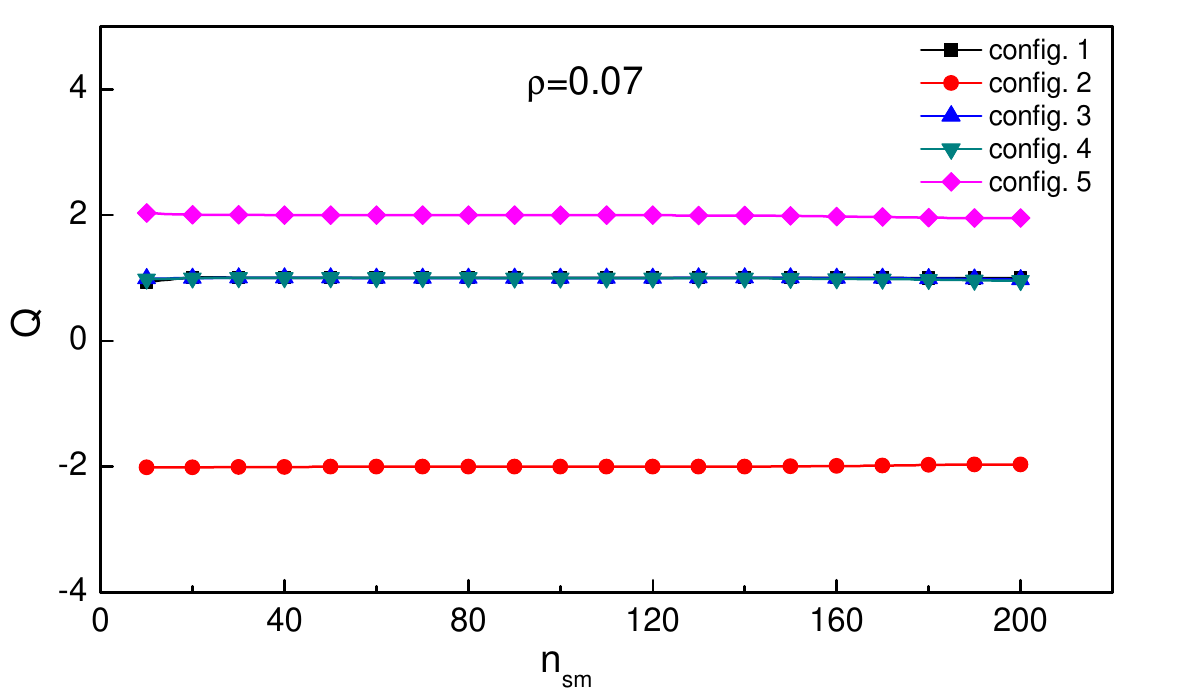}$~~~~$ \includegraphics[scale=0.5]{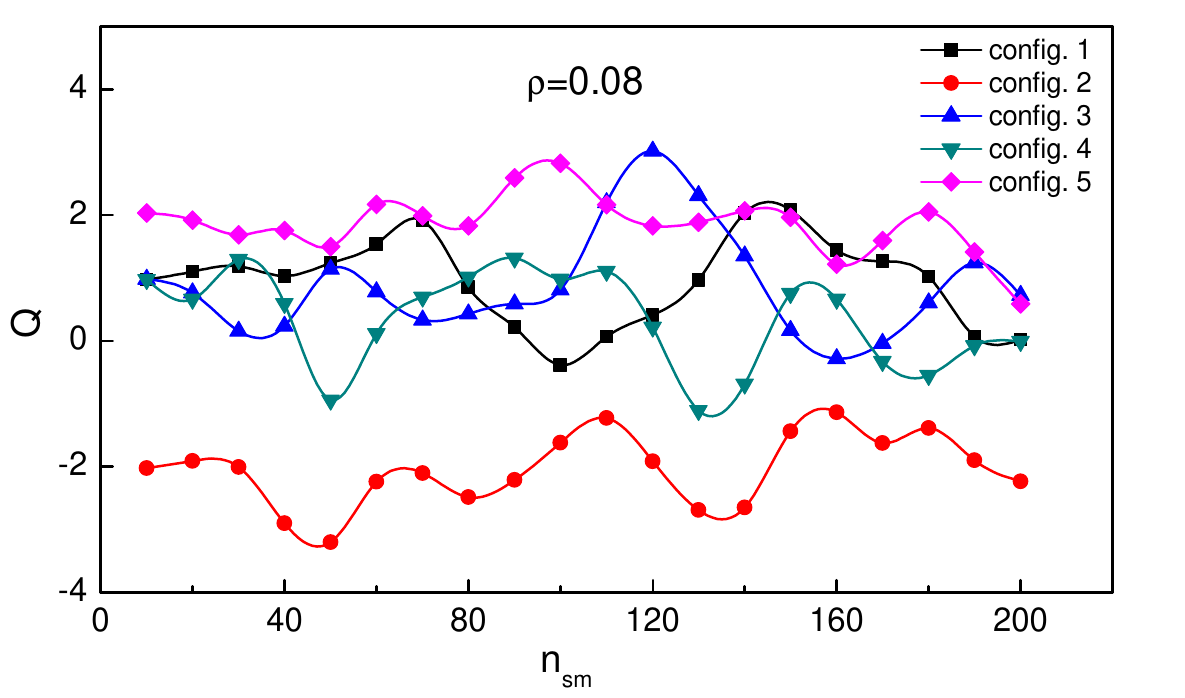}
\par\end{centering}
\caption{Smooth tests for different $\rho$ at $N_{t}=40$. Tests at other temperatures show similar situation, except that at lower temperatures, more steps are needed to extract the nearly integer values of $Q$.}
\label{fig1}
\end{figure}

There are three free parameters in the over-improved stout-link smearing method: the over-improved parameter $\epsilon$, the smearing parameter $\rho_{\mathrm{sm}}$, and $n_{\mathrm{sm}}$ -- the number of smoothing steps taken on each configuration. Moran and Leinweber had practically determined $\epsilon=-0.25$ to maximize the life of instantons under iterative smearing procedure, which we also kept fixed in our study. We check this on several configurations at different temperature and confirmed that topological charge approaches integral number quickly as long as $n_\mathrm{sm}\ge10\sim40$ in most situation, and keeps almost the same value for thousands of steps, which will be presented at the last of this sector.

For the other two parameters $\rho_{\mathrm{sm}}$ and $n_{\mathrm{sm}}$, we perform some tests to find suitable values for them. Five configurations are investigated for each temperature and smoothed 200 times for $\rho_{\mathrm{sm}}=0.01\sim0.08$, and the results of topological charge for some some typical values of $\rho_{\mathrm{sm}}$ are shown in Figure \ref{fig1}. It's obvious that the larger $\rho_\mathrm{sm}$ becomes, the faster the topological charge becomes an integer. However, in practice we found that when $\rho_{\mathrm{sm}}$ is equal to or larger than about 0.08, smoothing leads to completely unstable results, as shown in lower right panel in Figure \ref{fig1}, which is the same situation for all $N_t$. Finally we set $n_{\mathrm{sm}}=40$ and $\rho_{\mathrm{sm}}=0.05$ for all lattices (except for $N_{t}=128$ we set $\rho_{\mathrm{sm}}=0.07$, keeping $n_{\mathrm{sm}}=40$ ).

\begin{figure}[h t b]
\begin{centering}
{\includegraphics[scale=0.5]{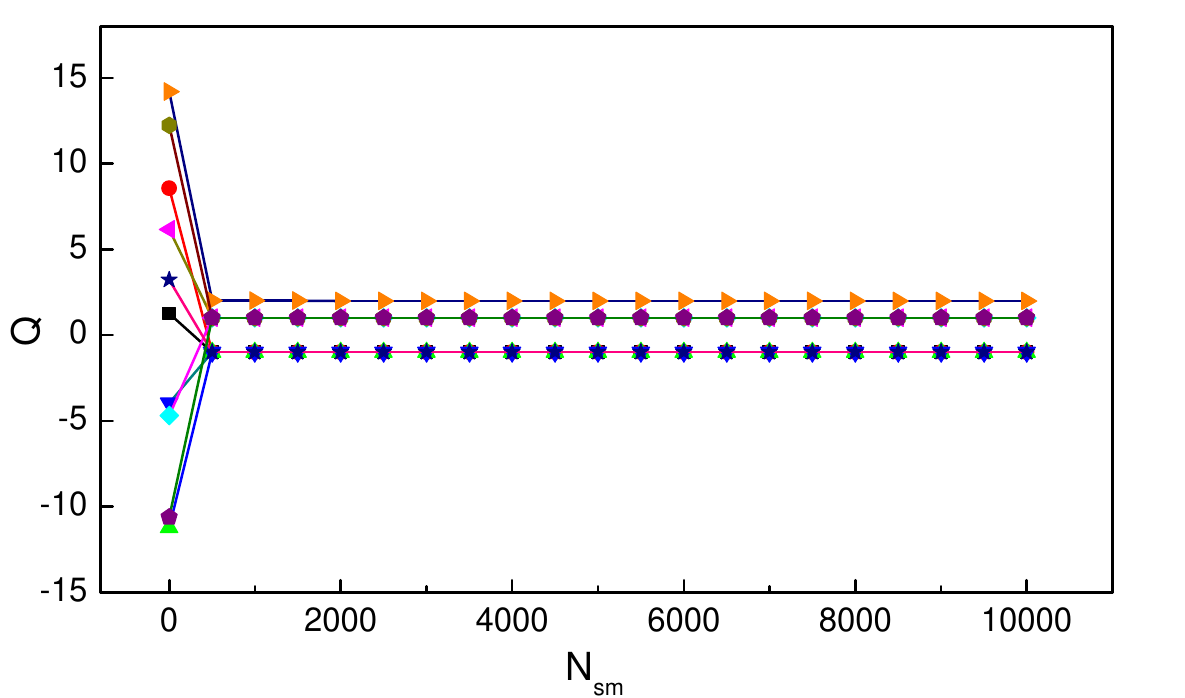} } $~~~~$ {\includegraphics[scale=0.5]{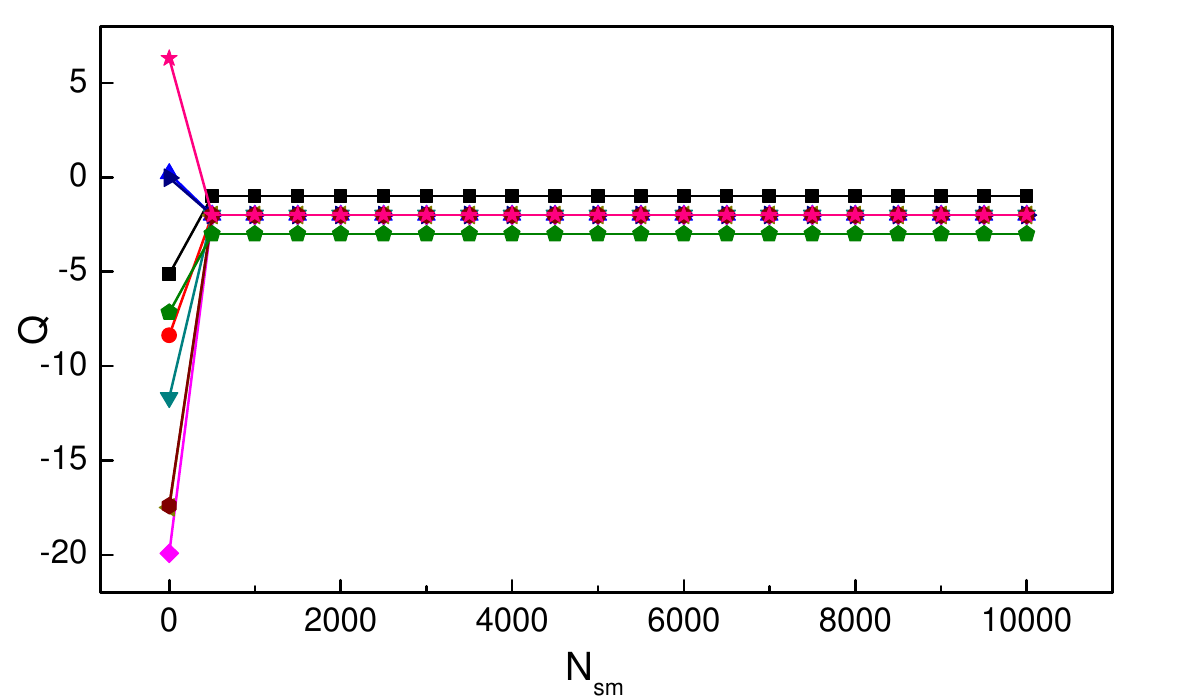} }
\par\end{centering}
\caption{Stability tests at $T=1.19T_c$ (left) and $T=0.79T_c$ (right) on 10 configurations with nonzero $Q$.}
\label{fig:smoothtest}
\end{figure}

By the way, we check the stability of the over-improved stout-link smearing method, which is supposed to be an effective and cheaper method (compared to the fermionic method) to suppressing the ultraviolet fluctuations on the lattices while preserving topological structures as long-lived as possible during smoothing. One of the most important problems of smoothing-type methods is the instability of topological charge when smoothing is carried out. We made several tests to check this, such as 10000 steps smearing on different configurations with nonzero topological charge $Q$ both below and above $T_c$. Part of the results are shown in Figure \ref{fig:smoothtest}. It indicates that, within the over-improved stout-link smearing method, with proper parameters and improved field strength tensor adopted, one can preserved instanton-like objects for long enough time without causing the system to fall into the trivial topological sector. Therefore, the result for the topological charge is robust and almost independent of the number of smoothing steps once an integral value for $Q$ is obtained for most of configurations. Notice that we measure the topological charge after every 500 steps of smoothing in stability tests while for most of configurations $Q$ is almost an integer after about 40 steps.

\section{Results and discussion}
We extract 1000 topological charge data for each temperature on anisotropic lattices and 5000 samples on $24^4$ isotropic lattice, which are all obtained after 40 steps of smoothing procedures. Most of these values are nearly integers. Those $Q$ values that deviate larger than 0.1 from an integer -- which we called "bad points" -- are less than $4\%$ for each lattice setting. The number of "bad points" grows slightly as the temperature approaches zero and for $N_{t}=128$ (corresponding to nearly zero temperature) we set $\rho_{\mathrm{sm}}=0.07$ to ensure that 40 steps of smooth are practically enough to avoid too much "bad points".
\begin{figure}[h t b]
\begin{centering}
{\includegraphics[scale=0.4]{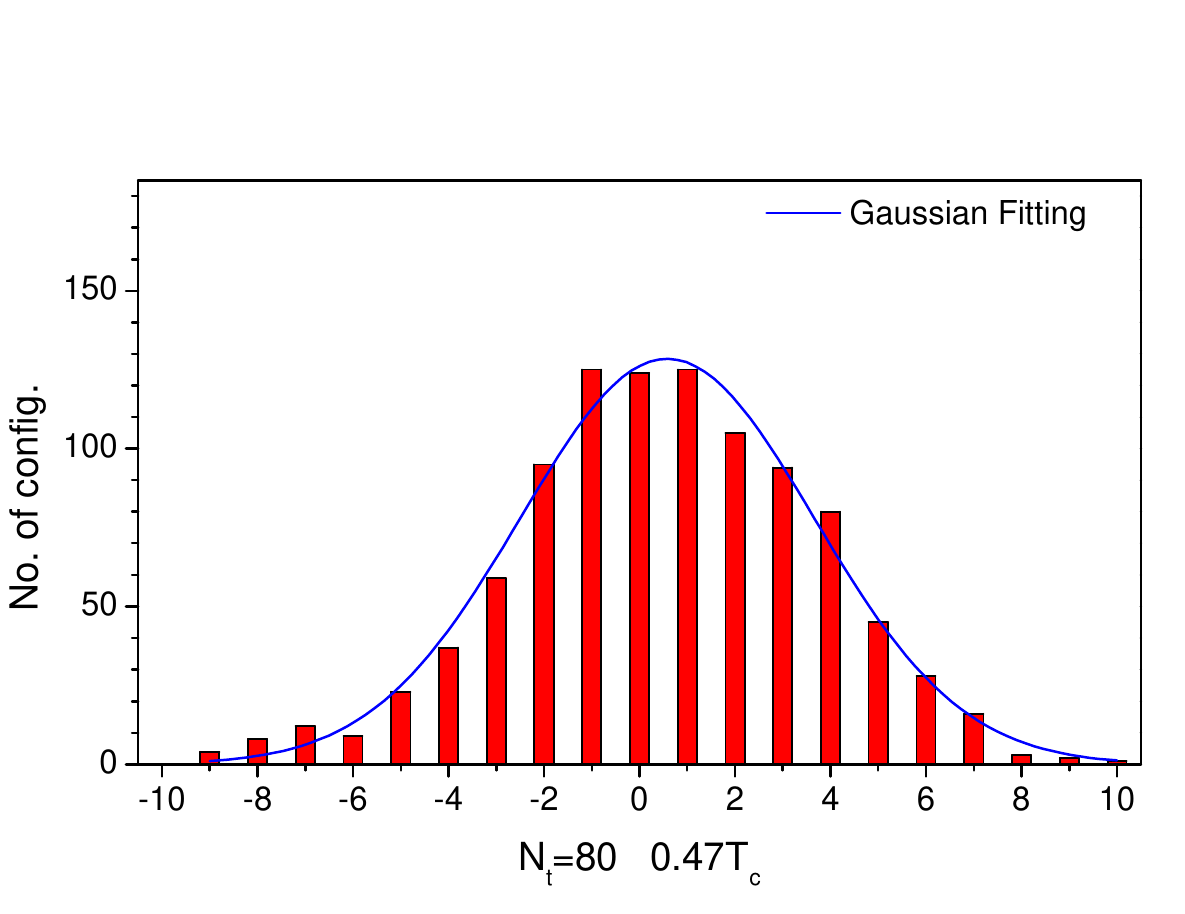} } $~~~~$ {\includegraphics[scale=0.4]{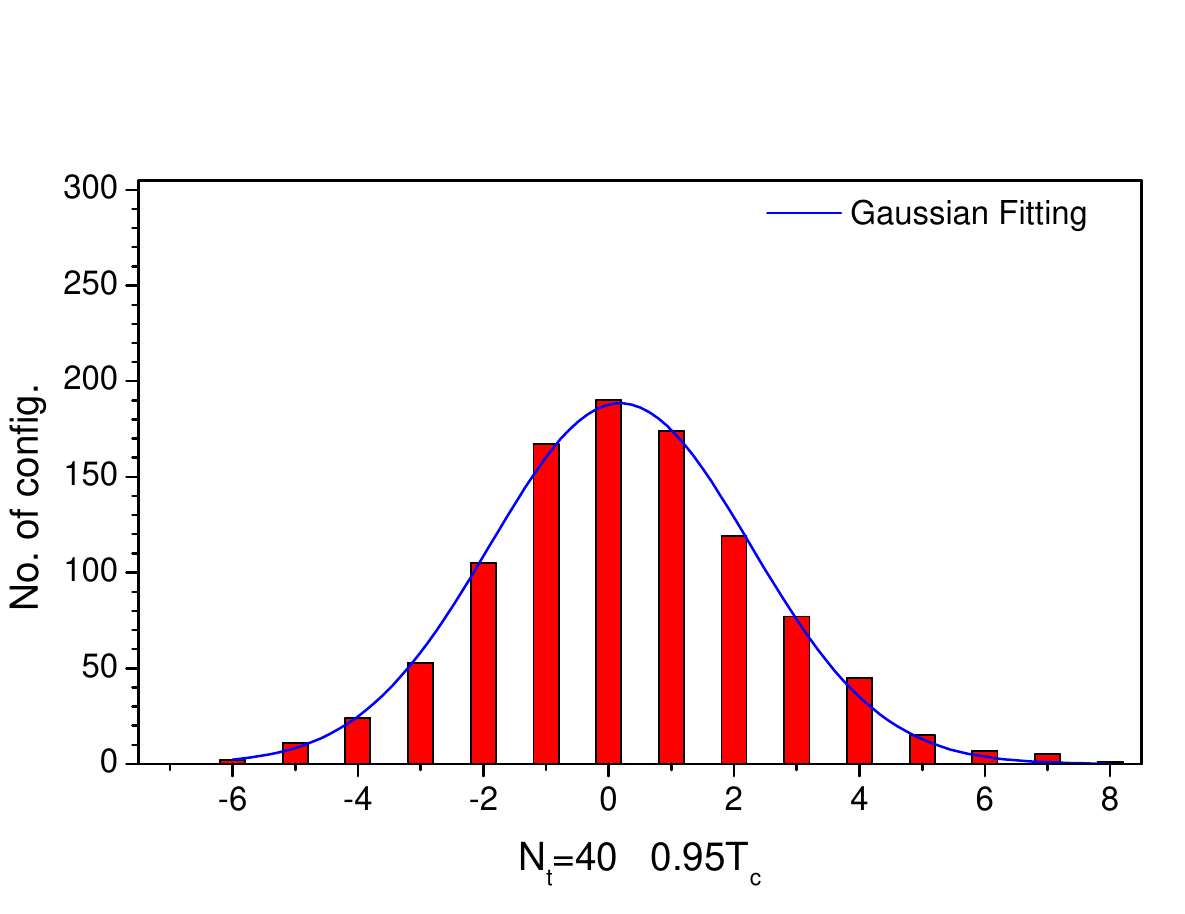} }
\par\end{centering}
\begin{centering}
{\includegraphics[scale=0.4]{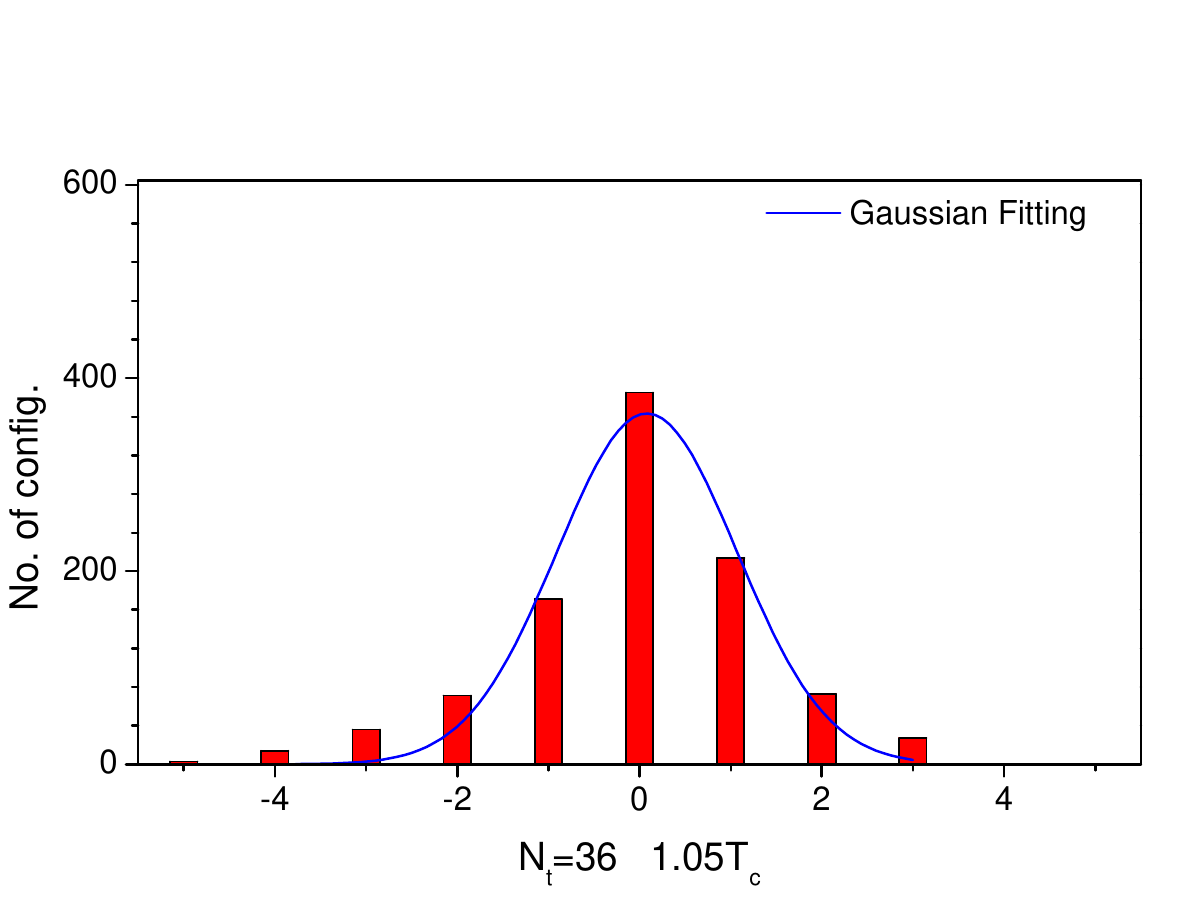} } $~~~~$ {\includegraphics[scale=0.4]{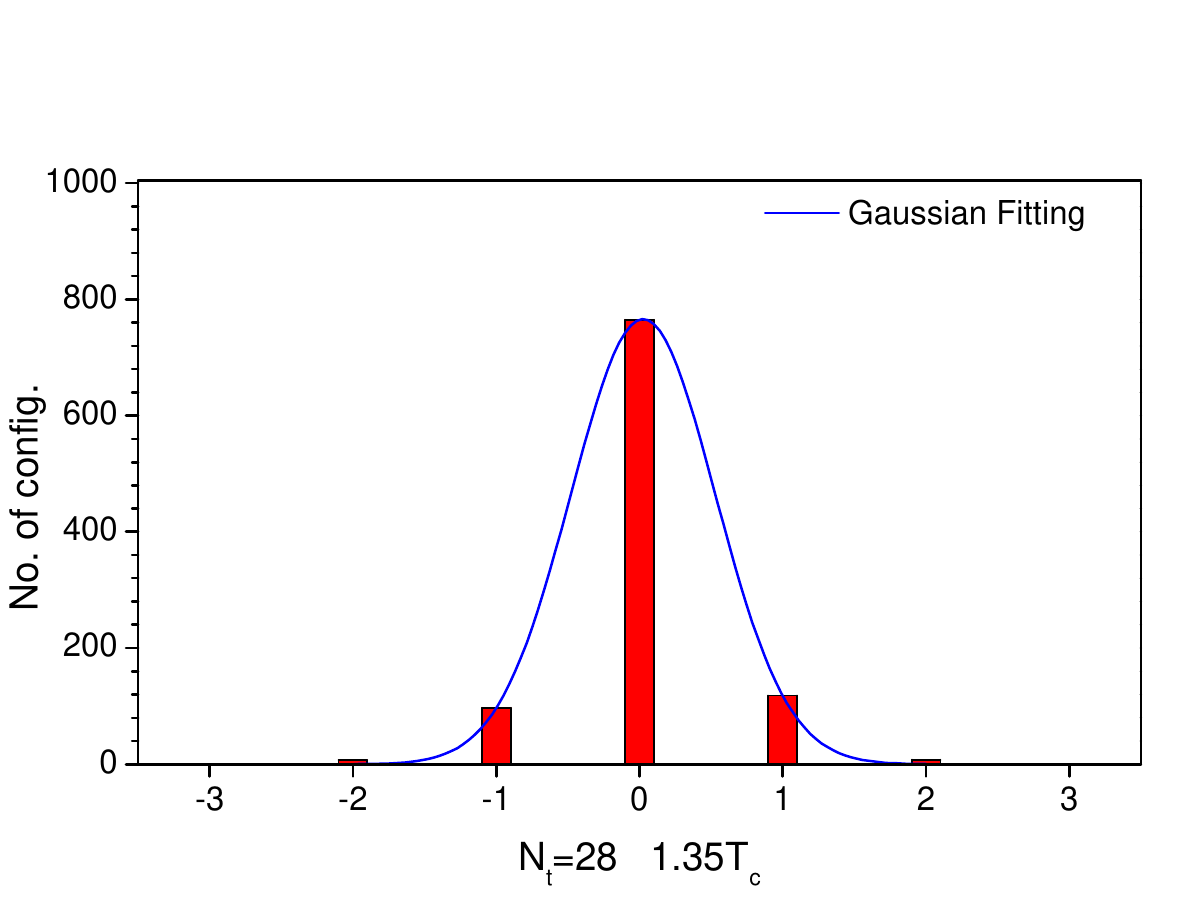} }
\par\end{centering}
\caption{Distribution of the topological charge at different temperature.}
\label{fig:Q_distri}
\end{figure}

Firstly, we display the histograms of the topological charge $Q$ for different temperatures in Figure \ref{fig:Q_distri}. As the temperature increases, distribution of $Q$ becomes narrower as expected, that the topological excitations are suppressed above $T_c$. We fit the histograms of $Q$ with a Gaussian function which are denoted as the blue lines in Figure \ref{fig:Q_distri}. The deviation from the Gaussian distribution comes from nonzero higher order cumulants and limited sampling.

\begin{figure}[h t b]
\begin{centering}
\includegraphics[scale=0.5]{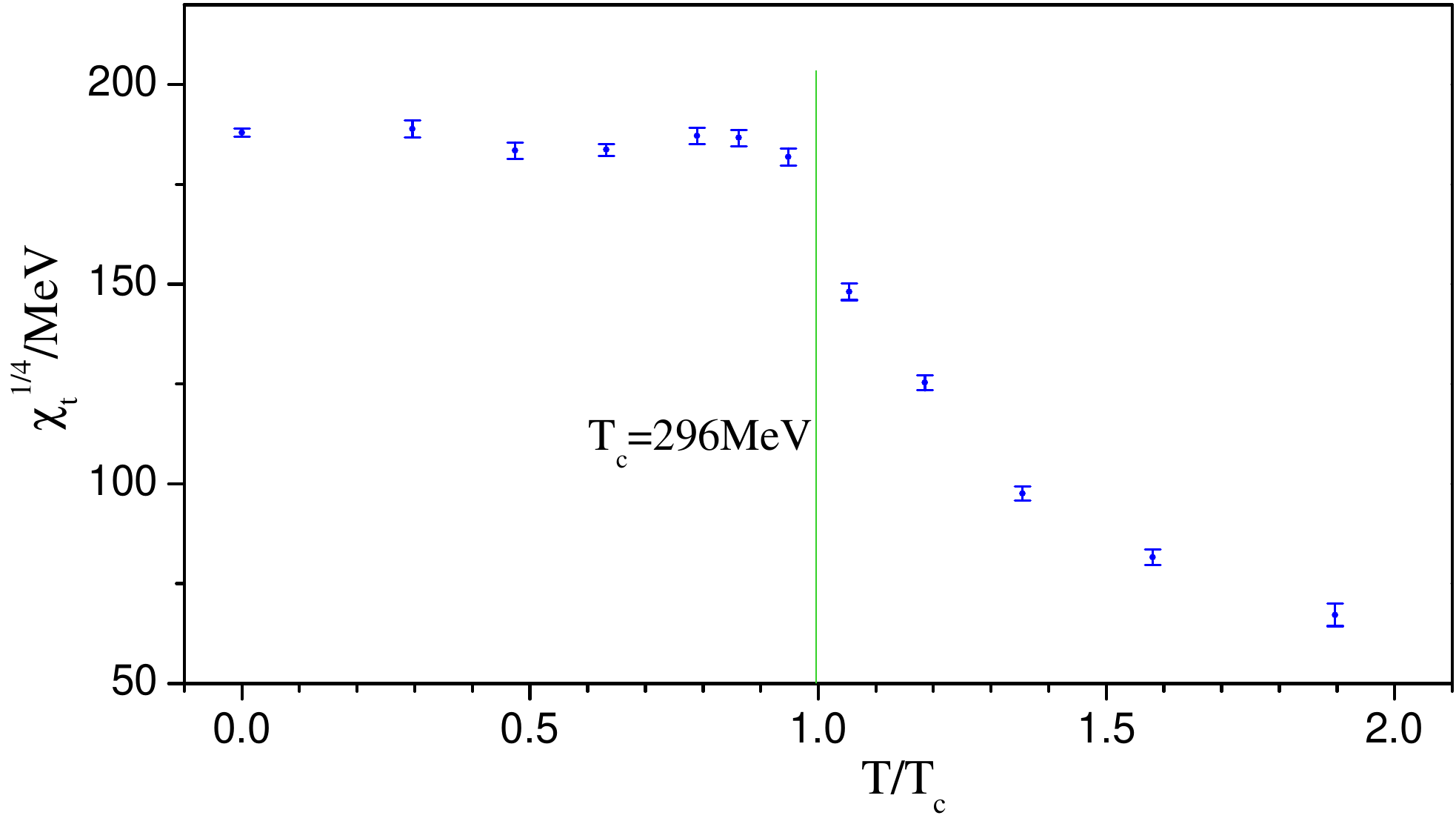}
\par\end{centering}
\caption{Results for $\chi^{1/4}_t$ vs. temperature}
\label{fig:chi result}
\end{figure}

\begin{figure}[h t b]
\begin{centering}
\includegraphics[scale=0.5]{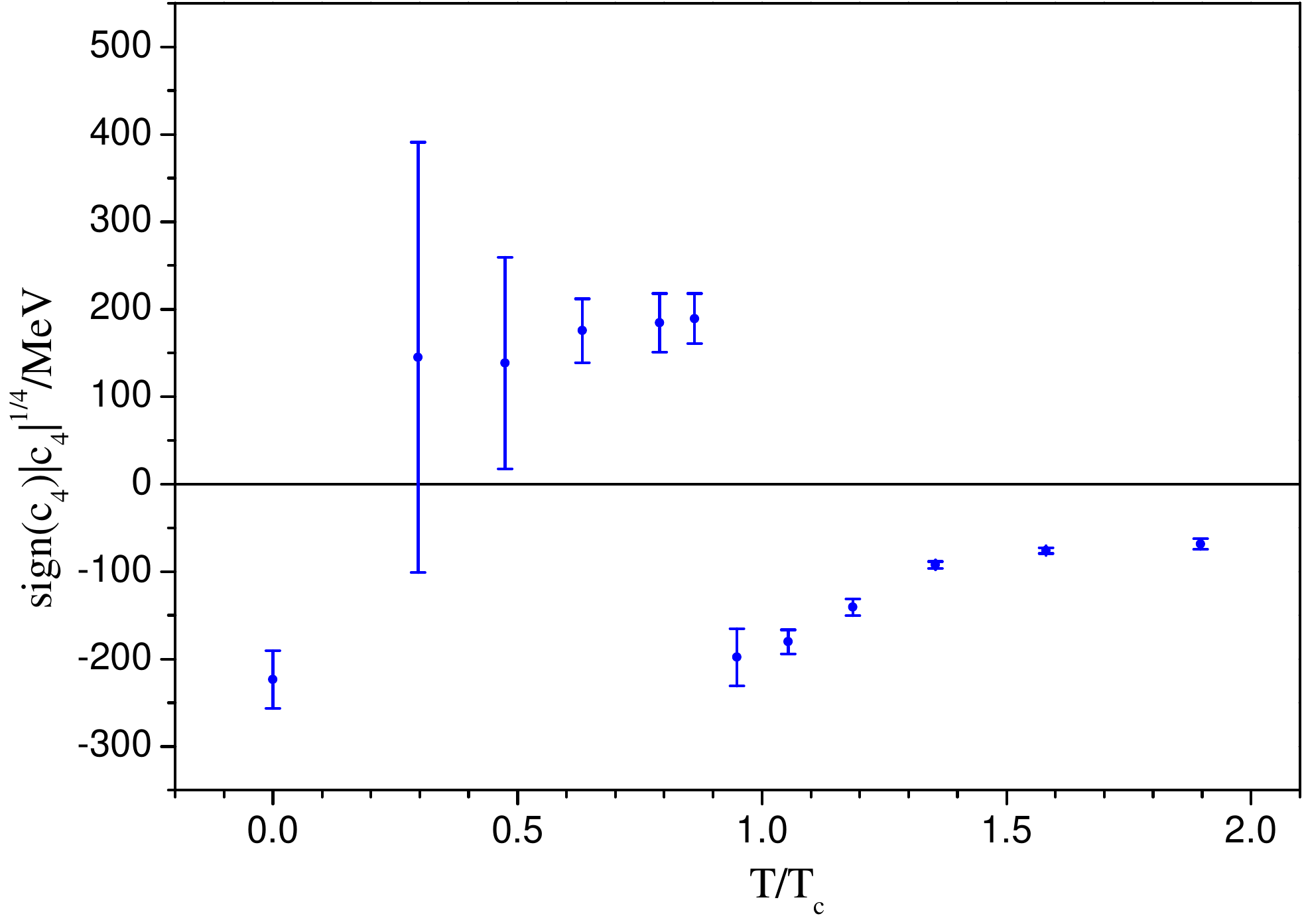}
\par\end{centering}

\caption{$c_{4}^{1/4}$ of $Q$. For convenient, we show $\left|c_{4}\right|^{1/4}$
with the same sign of $c_{4}$ }
\label{fig:c4}
\end{figure}

Next, the susceptibility $\chi_t$ and the 4th-order cumulant $c_4$ of $Q$ as well as the ratio $R$ are worked out at different temperatures, and the results are shown in Figure \ref{fig:chi result}, Figure \ref{fig:c4} and Table \ref{tab:chi result} where the errors are statistical and estimated by jackknife method.
These quantities from original topological charge data as well as $Q$ after rounding off are calculated, which make almost no differences. So the results presented are based on original $Q$ measured.

The susceptibility at $T=0$ reads $\chi_{t}=\left(188(1)\mathrm{MeV}\right)^4$ which is consistent with former results by other methods. It also clearly shows that topological excitation exists even at $T=1.9T_{c}$, where $\chi_t=\left(67\mathrm{MeV}\right)^4$ and $4.2\%$ configurations have nonzero topological charge. That means instanton-like structures may exist even above $T_c$ to nearly $2T_c$ supporting the conclusion of Gattringer {\em et al}. As Figure \ref{fig:chi result} shows, a crossover behavior for $\chi_{t}$ decreasing from $\left(188(1)\mathrm{MeV}\right)^{4}$ to $\left(67(3)\mathrm{MeV}\right)^{4}$ is found. For all configurations below $T_c$ (from 0 to 0.95$T_c$), $\chi_t$ stays around $(177\mathrm{MeV})^4\sim (189\mathrm{MeV})^4$. From $24^3\times36$ configurations corresponding to 1.05$T_c$, $\chi_t$ starts to decline as the temperature increases. Notice that Figure \ref{fig:chi result} shows the value of $\chi_t^{1/4}$, and $\chi_t$ decreases much faster than seen above $T_c$, consistent with results in Ref.~\cite{Gattringer2002}. For example, at $T=1.05T_c$ where $\chi_t=(148\mathrm{MeV})^4$ which is $38.4\%$ of its zero-temperature value, while at $T=1.9T_c$, $\chi_t=(67\mathrm{MeV})^4$ has dropped to about only $1.6\%$ of its zero-temperature value. The value of $\chi_t$ remains almost constant below $T_c$ and declines rather rapidly from $0.95T_c$ to $1.05T_c$ signaling for the phase transition.

As shown in Table \ref{tab:chi result}, $R$ approaches $-1$ from below in the interval {[}1.05$T_{c}$,1.19$T_{c}${]} while above $1.35T_{c}$ $R$ values not far away from $-1$, in agreement with the result of Ref.~\cite{Bonati2014}, in which $b_2=R/12$ approaches $-1/12$ from below as $T$ goes beyond $1.15T_c$ or so. This is also consistent with the value predicted by the diluted instanton gas model. 
Similar coincidence has also been found in a recent separate lattice study~\cite{Borsanyi:2015cka}. Besides, although the error bars grow larger as the temperature approaches zero, $c_4$ shows a step-like behavior near $T_c$ that could be seen in Figure \ref{fig:c4}. However, these behaviors couldn't be the proof of the existence of discontinuity between $0.86T_c$ and $0.95T_c$, that may be related to chiral and deconfinement phase transition, and should be carefully examined with more accurate calculations and finer temperature resolution. It is also found that, for pure gauge theories, the calculations of $c_{4}$ and $R$ do need much more samples to suppress the static errors, so we can't make any conclusion from the results of $c_{4}$ and $R$ near $T=0$. The statistical errors of $\chi_{t}$ and $c_{4}$ could be evaluated according to \cite{Giusti:2007tu}, which indeed shows much noisier signals of $c_{4}$ than those of $\chi_{t}$ near zero temperature.

\begin{table}[h t b]
\begin{centering}
\begin{tabular}{|c|c|c|c|c|c|}
\hline
$config.$ & $T/\mathrm{MeV}$ & $T/T_c$ & $\chi^{1/4}_t/\mathrm{MeV}$ & $c_4/{\mathrm{MeV}}^4$ & $R(c_4/\chi)$\tabularnewline
\hline
\hline
$24^3\times20$ & 562 & 1.90 & $67(3)$ & $-(69(6))^4$ & $-1.09(26)$\tabularnewline
\hline
$24^3\times24$ & 468 & 1.58 & $82(2)$ & $-(76(3))^4$ & $-0.75(10)$\tabularnewline
\hline
$24^3\times28$ & 401 & 1.35 & $98(2)$ & $-(93(4))^4$ & $-0.81(11)$\tabularnewline
\hline
$24^3\times32$ & 351 & 1.19 & $125(2)$ & $-(141(10))^4$ & $-1.58(36)$\tabularnewline
\hline
$24^3\times36$ & 312 & 1.05 & $148(2)$ & $-(180(14))^4$ & $-2.20(60)$\tabularnewline
\hline
$24^3\times40$ & 281 & 0.95 & $182(2)$ & $-(198(33))^4$ & $-1.38(84)$\tabularnewline
\hline
$24^3\times44$ & 255 & 0.86 & $187(2)$ & $(189(29))^4$ & $1.05(62)$\tabularnewline
\hline
$24^3\times48$ & 234 & 0.79 & $187(2)$ & $(184(33))^4$ & $0.94(68)$\tabularnewline
\hline
$24^3\times60$ & 187 & 0.63 & $184(1)$ & $(175(37))^4$ & $0.83(72)$\tabularnewline
\hline
$24^3\times80$ & 140 & $0.47$ & $183(2)$ & $(138(121))^4$ & $0.31(1.16)$\tabularnewline
\hline
$24^3\times128$&  88 & $0.30$ & $189(2)$ & $(145(246))^4$ & $0.32(2.78)$\tabularnewline
\hline
\hline
$24^3\times24$ &     0 &    0 & $188(1)$ & $-(223(33))^4$ & $-2.00(1.14)$\tabularnewline
\hline

\end{tabular}
\par\end{centering}
\caption{$\chi^{1/4}_t$, $c_4$ and $R$ at different temperatures. Last row is corresponding to isotropic configurations at $T=0$.}
\label{tab:chi result}
\end{table}

\section{Summary}
Topological susceptibility and the 4th order cumulant of topological charge in pure $\mathrm{SU}(3)$ gauge theory are calculated using anisotropic lattices from zero temperature to $1.9T_c$. The computation utilizes over-improved stout-link smearing method, employing 3-loop highly improved field strength tensor. The results obtained are consistent with than existing results in literatures using other methods, while extended to higher temperature. Once appropriate parameters are adopted, it is confirmed that the value of the topological charge is stable and robust during the smoothing procedure after becoming an integer. We also confirm the existence of topological excitations around $1.9T_c$. The asymptotic behavior of the ratio $c_4/\chi_t$ above $1.19T_c$ is consistent with the diluted instanton gas model. Furthermore, a step-like behavior of $c_4$ around $T_c$ inspires further investigation of high order cumulants of topological charge.

\section*{Acknowledgments}
This work was mainly run on Magic supercomputer in Shanghai Supercomputer Center and partially on cluster of IFTS at Zhejiang University as well as Tianhe-2 at NSCC in Guangzhou. This work is supported in part by the National Science Foundation of China (NSFC) under the project No.11335001, No.11275169, No.11075167, No.11105153. It is also supported in part by the DFG and the NSFC (No.11261130311) through funds provided to the Sino-German CRC 110 "Symmetries and the Emergence of Structure in QCD".
This work was also funded in part by National Basic Research Program of China (973 Program)
under code number 2015CB856700.

\section*{References}

\bibliographystyle{model1a-num-names}
\bibliography{refer}

\begin{thebibliography}{32}
\expandafter\ifx\csname natexlab\endcsname\relax\def\natexlab#1{#1}\fi
\providecommand{\url}[1]{\texttt{#1}}
\providecommand{\href}[2]{#2}
\providecommand{\path}[1]{#1}
\providecommand{\DOIprefix}{doi:}
\providecommand{\ArXivprefix}{arXiv:}
\providecommand{\URLprefix}{URL: }
\providecommand{\Pubmedprefix}{pmid:}
\providecommand{\doi}[1]{\href{http://dx.doi.org/#1}{\path{#1}}}
\providecommand{\Pubmed}[1]{\href{pmid:#1}{\path{#1}}}
\providecommand{\bibinfo}[2]{#2}
\ifx\xfnm\relax \def\xfnm[#1]{\unskip,\space#1}\fi
\bibitem[{Witten(1979)}]{witten1979current}
\bibinfo{author}{E.~Witten}, \bibinfo{journal}{Nuclear Physics B}
  \bibinfo{volume}{156} (\bibinfo{year}{1979}) \bibinfo{pages}{269--283}.
  \URLprefix
  \url{http://www.sciencedirect.com/science/article/pii/0550321379900312}.
  \DOIprefix\doi{10.1016/0550-3213(79)90031-2}.
\bibitem[{Veneziano(1979)}]{veneziano1979u}
\bibinfo{author}{G.~Veneziano}, \bibinfo{journal}{Nuclear Physics B}
  \bibinfo{volume}{159} (\bibinfo{year}{1979}) \bibinfo{pages}{213--224}.
  \URLprefix
  \url{http://www.sciencedirect.com/science/article/pii/0550321379903328}.
  \DOIprefix\doi{10.1016/0550-3213(79)90332-8}.
\bibitem[{Reichl(1980)}]{reichl1980modern}
\newblock {For example, see Page 143-144 in the following book:}
\bibinfo{author}{L.~E. Reichl}, \bibinfo{title}{A modern course in statistical
  physics}, \bibinfo{publisher}{University of Texas press Austin},
  \bibinfo{year}{1980}.
\bibitem[{Brower et~al.(2003)Brower, Chandrasekharan, Negele, and
  Wiese}]{Brower:2003yx}
\bibinfo{author}{R.~Brower}, \bibinfo{author}{S.~Chandrasekharan},
  \bibinfo{author}{J.~W. Negele}, \bibinfo{author}{U.~J. Wiese},
  \bibinfo{journal}{Phys. Lett.B} \bibinfo{volume}{B560} (\bibinfo{year}{2003})
  \bibinfo{pages}{64--74}. \DOIprefix\doi{10.1016/S0370-2693(03)00369-1}.
  \href{http://arxiv.org/abs/hep-lat/0302005}{\tt arXiv:hep-lat/0302005}.
\bibitem[{Aoki et~al.(2007)Aoki, Fukaya, Hashimoto, and Onogi}]{Aoki:2007ka}
\bibinfo{author}{S.~Aoki}, \bibinfo{author}{H.~Fukaya},
  \bibinfo{author}{S.~Hashimoto}, \bibinfo{author}{T.~Onogi},
  \bibinfo{journal}{Phys. Rev.D} \bibinfo{volume}{D76} (\bibinfo{year}{2007})
  \bibinfo{pages}{054508}. \DOIprefix\doi{10.1103/PhysRevD.76.054508}.
  \href{http://arxiv.org/abs/0707.0396}{\tt arXiv:0707.0396}.
\bibitem[{Giusti et~al.(2007)Giusti, Petrarca, and Taglienti}]{Giusti:2007tu}
\bibinfo{author}{L.~Giusti}, \bibinfo{author}{S.~Petrarca},
  \bibinfo{author}{B.~Taglienti}, \bibinfo{journal}{Phys. Rev.D}
  \bibinfo{volume}{D76} (\bibinfo{year}{2007}) \bibinfo{pages}{094510}.
  \DOIprefix\doi{10.1103/PhysRevD.76.094510}.
  \href{http://arxiv.org/abs/0705.2352}{\tt arXiv:0705.2352}.
\bibitem[{C{\`e} et~al.(2015)C{\`e}, Consonni, Engel, and Giusti}]{Ce:2015qha}
\bibinfo{author}{M.~C{\`e}}, \bibinfo{author}{C.~Consonni},
  \bibinfo{author}{G.~P. Engel}, \bibinfo{author}{L.~Giusti}
  (\bibinfo{year}{2015}). \href{http://arxiv.org/abs/1506.06052}{\tt
  arXiv:1506.06052}.
\bibitem[{Petreczky(2012)}]{Petreczky2012}
\bibinfo{author}{P.~Petreczky}, \bibinfo{journal}{arXiv} \bibinfo{volume}{1}
  (\bibinfo{year}{2012}) \bibinfo{pages}{48}. \URLprefix
  \url{http://arxiv.org/abs/1203.5320}.
  \href{http://arxiv.org/abs/[hep-lat/1203.5320]}{\tt
  arXiv:[hep-lat/1203.5320]}.
\bibitem[{Del~Debbio et~al.(2005)Del~Debbio, Giusti, and Pica}]{DelDebbio2005}
\bibinfo{author}{L.~Del~Debbio}, \bibinfo{author}{L.~Giusti},
  \bibinfo{author}{C.~Pica}, \bibinfo{journal}{Phys. Rev. Lett.}
  \bibinfo{volume}{94} (\bibinfo{year}{2005}) \bibinfo{pages}{032003}.
  \URLprefix \url{http://link.aps.org/doi/10.1103/PhysRevLett.94.032003}.
  \DOIprefix\doi{10.1103/PhysRevLett.94.032003}.
\bibitem[{Alles et~al.(1997)Alles, D'Elia, and Di~Giacomo}]{Alles:1996nm}
\bibinfo{author}{B.~Alles}, \bibinfo{author}{M.~D'Elia},
  \bibinfo{author}{A.~Di~Giacomo}, \bibinfo{journal}{Nucl.Phys.}
  \bibinfo{volume}{B494} (\bibinfo{year}{1997}) \bibinfo{pages}{281--292}.
  \DOIprefix\doi{10.1016/S0550-3213(97)00205-8}.
  \href{http://arxiv.org/abs/hep-lat/9605013}{\tt arXiv:hep-lat/9605013}.
\bibitem[{de~Forcrand et~al.(1997)de~Forcrand, Garcia~Perez, Hetrick, and
  Stamatescu}]{deForcrand:1997ut}
\bibinfo{author}{P.~de~Forcrand}, \bibinfo{author}{M.~Garcia~Perez},
  \bibinfo{author}{J.~E. Hetrick}, \bibinfo{author}{I.-O. Stamatescu}
  (\bibinfo{year}{1997}). \href{http://arxiv.org/abs/[hep-lat/9802017]}{\tt
  arXiv:[hep-lat/9802017]}.
\bibitem[{Gattringer and Hoffmann(2002)}]{Gattringer2002}
\bibinfo{author}{C.~Gattringer}, \bibinfo{author}{R.~Hoffmann},
  \bibinfo{journal}{Physics Letters B} \bibinfo{volume}{535}
  (\bibinfo{year}{2002}) \bibinfo{pages}{358--362}. \URLprefix
  \url{http://www.sciencedirect.com/science/article/pii/S0370269302017574}.
  \DOIprefix\doi{10.1016/S0370-2693(02)01757-4}.
\bibitem[{Del~Debbio et~al.(2004)Del~Debbio, Panagopoulos, and
  Vicari}]{DelDebbio:2004rw}
\bibinfo{author}{L.~Del~Debbio}, \bibinfo{author}{H.~Panagopoulos},
  \bibinfo{author}{E.~Vicari}, \bibinfo{journal}{JHEP} \bibinfo{volume}{0409}
  (\bibinfo{year}{2004}) \bibinfo{pages}{028}.
  \DOIprefix\doi{10.1088/1126-6708/2004/09/028}.
  \href{http://arxiv.org/abs/hep-th/0407068}{\tt arXiv:hep-th/0407068}.
\bibitem[{L{\"u}scher(2010)}]{Luscher:2010iy}
\bibinfo{author}{M.~L{\"u}scher}, \bibinfo{journal}{JHEP} \bibinfo{volume}{08}
  (\bibinfo{year}{2010}) \bibinfo{pages}{071}.
  \DOIprefix\doi{10.1007/JHEP08(2010)071, 10.1007/JHEP03(2014)092}.
  \href{http://arxiv.org/abs/1006.4518}{\tt arXiv:1006.4518},
  \bibinfo{note}{[Erratum: JHEP03,092(2014)]}.
\bibitem[{Bonati et~al.(2013)Bonati, D'Elia, Panagopoulos, and
  Vicari}]{Bonati:2013tt}
\bibinfo{author}{C.~Bonati}, \bibinfo{author}{M.~D'Elia},
  \bibinfo{author}{H.~Panagopoulos}, \bibinfo{author}{E.~Vicari},
  \bibinfo{journal}{Phys. Rev. Lett.} \bibinfo{volume}{110}
  (\bibinfo{year}{2013}) \bibinfo{pages}{252003}.
  \DOIprefix\doi{10.1103/PhysRevLett.110.252003}.
  \href{http://arxiv.org/abs/1301.7640}{\tt arXiv:1301.7640}.
\bibitem[{Bonati and D'Elia(2014)}]{Bonati:2014tqa}
\bibinfo{author}{C.~Bonati}, \bibinfo{author}{M.~D'Elia},
  \bibinfo{journal}{Phys. Rev.D} \bibinfo{volume}{D89} (\bibinfo{year}{2014})
  \bibinfo{pages}{105005}. \DOIprefix\doi{10.1103/PhysRevD.89.105005}.
  \href{http://arxiv.org/abs/1401.2441}{\tt arXiv:1401.2441}.
\bibitem[{Moran and Leinweber(2008)}]{Moran2008}
\bibinfo{author}{P.~Moran}, \bibinfo{author}{D.~Leinweber},
  \bibinfo{journal}{Physical Review D} \bibinfo{volume}{77}
  (\bibinfo{year}{2008}) \bibinfo{pages}{094501}. \URLprefix
  \url{http://link.aps.org/doi/10.1103/PhysRevD.77.094501}.
  \DOIprefix\doi{10.1103/PhysRevD.77.094501}.
\bibitem[{Cossu et~al.(2011)Cossu, Aoki, Hashimoto, Kaneko, Matsufuru, Noaki,
  and Shintani}]{Cossu:2012gm}
\bibinfo{author}{G.~Cossu}, \bibinfo{author}{S.~Aoki},
  \bibinfo{author}{S.~Hashimoto}, \bibinfo{author}{T.~Kaneko},
  \bibinfo{author}{H.~Matsufuru}, \bibinfo{author}{J.-i. Noaki},
  \bibinfo{author}{E.~Shintani}, \bibinfo{journal}{PoS}
  \bibinfo{volume}{LATTICE2011} (\bibinfo{year}{2011}) \bibinfo{pages}{188}.
  \href{http://arxiv.org/abs/1204.4519}{\tt arXiv:1204.4519}.
\bibitem[{Bornyakov et~al.(2013)Bornyakov, Ilgenfritz, Martemyanov,
  Mitrjushkin, and M{\"u}ller-Preussker}]{Bornyakov:2013iva}
\bibinfo{author}{V.~Bornyakov}, \bibinfo{author}{E.-M. Ilgenfritz},
  \bibinfo{author}{B.~Martemyanov}, \bibinfo{author}{V.~Mitrjushkin},
  \bibinfo{author}{M.~M{\"u}ller-Preussker}, \bibinfo{journal}{Phys.Rev.D}
  \bibinfo{volume}{D87} (\bibinfo{year}{2013}) \bibinfo{pages}{114508}.
  \DOIprefix\doi{10.1103/PhysRevD.87.114508}.
  \href{http://arxiv.org/abs/1304.0935}{\tt arXiv:1304.0935}.
\bibitem[{Ilgenfritz et~al.(2014)Ilgenfritz, Martemyanov, and
  M{\"u}ller-Preussker}]{Ilgenfritz:2013oda}
\bibinfo{author}{E.-M. Ilgenfritz}, \bibinfo{author}{B.~Martemyanov},
  \bibinfo{author}{M.~M{\"u}ller-Preussker}, \bibinfo{journal}{Phys.Rev.D}
  \bibinfo{volume}{D89} (\bibinfo{year}{2014}) \bibinfo{pages}{054503}.
  \DOIprefix\doi{10.1103/PhysRevD.89.054503}.
  \href{http://arxiv.org/abs/1309.7850}{\tt arXiv:1309.7850}.
\bibitem[{Gross et~al.(1981)Gross, Pisarski, and Yaffe}]{Gross:1980br}
\bibinfo{author}{D.~J. Gross}, \bibinfo{author}{R.~D. Pisarski},
  \bibinfo{author}{L.~G. Yaffe}, \bibinfo{journal}{Rev. Mod. Phys.}
  \bibinfo{volume}{53} (\bibinfo{year}{1981}) \bibinfo{pages}{43}.
  \DOIprefix\doi{10.1103/RevModPhys.53.43}.
\bibitem[{Bilson-Thompson et~al.(2003)Bilson-Thompson, Leinweber, and
  Williams}]{BilsonThompson:2002jk}
\bibinfo{author}{S.~O. Bilson-Thompson}, \bibinfo{author}{D.~B. Leinweber},
  \bibinfo{author}{A.~G. Williams}, \bibinfo{journal}{Annals of Physics}
  \bibinfo{volume}{304} (\bibinfo{year}{2003}) \bibinfo{pages}{1--21}.
\bibitem[{Meng et~al.(2009)Meng, Li, Zhang, Chen, Liu, Liu, Ma, and
  Zhang}]{Meng2009}
\bibinfo{author}{X.~Meng}, \bibinfo{author}{G.~Li}, \bibinfo{author}{Y.~Zhang},
  \bibinfo{author}{Y.~Chen}, \bibinfo{author}{C.~Liu},
  \bibinfo{author}{Y.~Liu}, \bibinfo{author}{J.~Ma},
  \bibinfo{author}{J.~Zhang}, \bibinfo{journal}{Physical Review D}
  \bibinfo{volume}{80} (\bibinfo{year}{2009}) \bibinfo{pages}{114502}.
  \URLprefix \url{http://link.aps.org/doi/10.1103/PhysRevD.80.114502}.
  \DOIprefix\doi{10.1103/PhysRevD.80.114502}.
\bibitem[{Morningstar and Peardon(1997)}]{Morningstar:1997ff}
\bibinfo{author}{C.~J. Morningstar}, \bibinfo{author}{M.~J. Peardon},
  \bibinfo{journal}{Phys. Rev.} \bibinfo{volume}{D56} (\bibinfo{year}{1997})
  \bibinfo{pages}{4043--4061}. \DOIprefix\doi{10.1103/PhysRevD.56.4043}.
  \href{http://arxiv.org/abs/hep-lat/9704011}{\tt arXiv:hep-lat/9704011}.
\bibitem[{Morningstar and Peardon(1999)}]{Morningstar:1999rf}
\bibinfo{author}{C.~J. Morningstar}, \bibinfo{author}{M.~J. Peardon},
  \bibinfo{journal}{Phys. Rev.D} \bibinfo{volume}{D60} (\bibinfo{year}{1999})
  \bibinfo{pages}{034509}. \DOIprefix\doi{10.1103/PhysRevD.60.034509}.
  \href{http://arxiv.org/abs/hep-lat/9901004}{\tt arXiv:hep-lat/9901004}.
\bibitem[{Chen et~al.(2006)Chen, Alexandru, Dong, Draper, Horv\'ath, Lee, Liu,
  Mathur, Morningstar, Peardon, Tamhankar, Young, and
  Zhang}]{PhysRevD.73.014516}
\bibinfo{author}{Y.~Chen}, \bibinfo{author}{A.~Alexandru},
  \bibinfo{author}{S.~J. Dong}, \bibinfo{author}{T.~Draper},
  \bibinfo{author}{I.~Horv\'ath}, \bibinfo{author}{F.~X. Lee},
  \bibinfo{author}{K.~F. Liu}, \bibinfo{author}{N.~Mathur},
  \bibinfo{author}{C.~Morningstar}, \bibinfo{author}{M.~Peardon},
  \bibinfo{author}{S.~Tamhankar}, \bibinfo{author}{B.~L. Young},
  \bibinfo{author}{J.~B. Zhang}, \bibinfo{journal}{Phys. Rev. D}
  \bibinfo{volume}{73} (\bibinfo{year}{2006}) \bibinfo{pages}{014516}.
  \URLprefix \url{http://link.aps.org/doi/10.1103/PhysRevD.73.014516}.
  \DOIprefix\doi{10.1103/PhysRevD.73.014516}.
\bibitem[{Liu et~al.(2006)Liu, Chen, Gong, Li, Liu, and Meng}]{Liu:2006dp}
\bibinfo{author}{W.~Liu}, \bibinfo{author}{Y.~Chen}, \bibinfo{author}{M.~Gong},
  \bibinfo{author}{X.~Li}, \bibinfo{author}{C.~Liu}, \bibinfo{author}{G.-z.
  Meng}, \bibinfo{journal}{Mod. Phys. Lett.} \bibinfo{volume}{A21}
  (\bibinfo{year}{2006}) \bibinfo{pages}{2313--2322}.
  \DOIprefix\doi{10.1142/S021773230601989X}.
  \href{http://arxiv.org/abs/hep-lat/0603015}{\tt arXiv:hep-lat/0603015}.
\bibitem[{Borsanyi et~al.(2012)Borsanyi, Durr, Fodor, Katz, Krieg, Kurth,
  Mages, Schafer, and Szabo}]{Borsanyi:2012zr}
\bibinfo{author}{S.~Borsanyi}, \bibinfo{author}{S.~Durr},
  \bibinfo{author}{Z.~Fodor}, \bibinfo{author}{S.~D. Katz},
  \bibinfo{author}{S.~Krieg}, \bibinfo{author}{T.~Kurth},
  \bibinfo{author}{S.~Mages}, \bibinfo{author}{A.~Schafer},
  \bibinfo{author}{K.~K. Szabo}  (\bibinfo{year}{2012}).
  \href{http://arxiv.org/abs/1205.0781}{\tt arXiv:1205.0781}.
\bibitem[{Belavin et~al.(1975)Belavin, Polyakov, Schwartz, and
  Tyupkin}]{belavin1975pseudoparticle}
\bibinfo{author}{A.~Belavin}, \bibinfo{author}{A.~Polyakov},
  \bibinfo{author}{A.~Schwartz}, \bibinfo{author}{Y.~Tyupkin},
  \bibinfo{journal}{Physics Letters B} \bibinfo{volume}{59}
  (\bibinfo{year}{1975}) \bibinfo{pages}{85 -- 87}. \URLprefix
  \url{http://www.sciencedirect.com/science/article/pii/037026937590163X}.
  \DOIprefix\doi{10.1016/0370-2693(75)90163-X}.
\bibitem[{L{\"u}scher(1982)}]{luscher1982topology}
\bibinfo{author}{M.~L{\"u}scher}, \bibinfo{journal}{Communications in
  Mathematical Physics} \bibinfo{volume}{85} (\bibinfo{year}{1982})
  \bibinfo{pages}{39--48}. \URLprefix
  \url{http://dx.doi.org/10.1007/BF02029132},
  \bibinfo{note}{10.1007/BF02029132}.
\bibitem[{Garcia~Perez et~al.(1994)Garcia~Perez, Gonzalez-Arroyo, Snippe, and
  van Baal}]{GarciaPerez:1993ki}
\bibinfo{author}{M.~Garcia~Perez}, \bibinfo{author}{A.~Gonzalez-Arroyo},
  \bibinfo{author}{J.~R. Snippe}, \bibinfo{author}{P.~van Baal},
  \bibinfo{journal}{Nucl. Phys.} \bibinfo{volume}{B413} (\bibinfo{year}{1994})
  \bibinfo{pages}{535--552}. \DOIprefix\doi{10.1016/0550-3213(94)90631-9}.
  \href{http://arxiv.org/abs/hep-lat/9309009}{\tt arXiv:hep-lat/9309009}.
\bibitem[{Bonati et~al.(2014)Bonati, D'Elia, Panagopoulos, and
  Vicari}]{Bonati2014}
\bibinfo{author}{C.~Bonati}, \bibinfo{author}{M.~D'Elia},
  \bibinfo{author}{H.~Panagopoulos}, \bibinfo{author}{E.~Vicari},
  \bibinfo{journal}{PoS} \bibinfo{volume}{LATTICE2013} (\bibinfo{year}{2014})
  \bibinfo{pages}{136}. \href{http://arxiv.org/abs/1309.6059}{\tt
  arXiv:1309.6059}.
\bibitem[{S.~Borsanyi et~al.(2014)}]{Borsanyi:2015cka}
\bibinfo{author}{S.~Borsanyi, {\it et al.}},
  \href{http://arxiv.org/abs/1508.06917}{\tt
  arXiv:1508.06917}.
\end{thebibliography}

\end{document}